\documentclass[letterpaper, 11pt]{article}
\usepackage{epsfig}
\usepackage{graphicx}
\usepackage{amsmath}
\usepackage{float}

\usepackage{authblk}

\begin{document}

\title{\bf Big Bang nucleosynthesis revisited via Trojan Horse Method  measurements}

\author{R.G. Pizzone$^{1}$, R. Spart\'a$^{1,2}$, C.A. Bertulani$^{3}$, C. Spitaleri$^{1,2}$,  M. La Cognata$^{1}$, J.  Lalmansingh$^{3}$, L. Lamia$^{2}$, A. Mukhamedzhanov$^{4}$, A. Tumino$^{1,5}$}

\affil{\it $^1$INFN - Laboratori Nazionali del Sud, Catania, Italy \\
\it $^2$Dipartimento di Fisica e Astronomia, Universit\`a degli Studi di Catania, Catania, Italy\\
\it $^3$Department of Physics and Astronomy, Texas A\&M University-Commerce, Commerce, TX 75025, USA \\
\it $^4$ Cyclotron Institute, Texas A\& M University, College Station, TX, USA\\
\it $^5$ Universit\`a degli Studi di Enna ``Kore", Enna, Italy
}

\maketitle

\begin{abstract}
Nuclear reaction rates are among the most important input for  understanding the primordial nucleosynthesis and therefore for a quantitative description of the early Universe. An up-to-date compilation of direct cross sections of $^{2}$H(d,p)$^{3}$H, $^{2}$H(d,n)$^{3}$He, $^7$Li(p,$\alpha$)$^4$He  and $^{3}$He(d,p)$^{4}$He reactions is given. These are among the most uncertain cross sections used and input for Big Bang nucleosynthesis calculations. Their measurements through the Trojan Horse Method (THM) are also reviewed and compared with direct data.
The reaction rates and the corresponding  recommended  errors in this work were used as input for primordial nucleosynthesis calculations to evaluate their impact on the $^2$H, $^{3,4}$He and $^7$Li primordial abundances, which are then compared with  observations. 
\end{abstract}


\maketitle

\section{Introduction}

Over the last decades Big Bang Nucleosynthesis (BBN) has emerged as one of the pillars of the Big Bang model, together with the Hubble expansion and the Cosmic Microwave Background (CMB) radiation \cite{Stei07}. BBN probes the Universe to the earliest times, from a fraction of second to few minutes. 
It involves events that occurred at temperatures below 1 MeV, and naturally plays a key role in forging the connection between cosmology and nuclear physics \cite{OSW00,Fie06}. Focusing only on the products of the BBN, according to the Standard Big Bang Nucleosynthesis model (SBBN), only the formation of light nuclei ($^2$H,$^{3,4}$He,$^{7}$Li) is predicted in observable quantities, starting from protons and neutrons. Today, with the only exception of $^3$He and lithium, the abundances of these isotopes in the appropriate astrophysical environments are rather consistent with SBBN predictions \cite{Isr12}. A comparison between the primordial abundances from WMAP observations and the calculated ones constrains the baryon-to-photon ratio, $\eta$, which is the only free parameter in the presently accepted model of the SBBN. A recent observation yields $\eta =6.16 \pm 0.15\times 10^{-10}$ \cite{Kom11}, which is the value that we adopt in our calculations.

BBN nucleosynthesis requires several nuclear physics inputs and, among them, an important role is played by nuclear reaction rates. Due to the relatively small amount of key nuclear species involved in the BBN nuclear reaction network, only 12 reactions play a major role \cite{KT90}. They are listed in Table \ref{rent}.

\begin{table}[htbp]
\begin{center}
\vspace{0.0cm}
\begin{tabular}{|l|l|l|l|l|}
\hline
\hline
${\rm n} \leftrightarrow {\rm p}$ \hfil {\bf (1)} &${\rm p(n,}\gamma{\rm )d}$\hfil {\bf (2)}   & ${\rm d(p,}\gamma)^3{\rm He}$\hfil {\bf (3)}  & ${\rm d(d, p)t}	^{\bf (*)}$\hfil  {\bf (4)}\\ \hline\hline
${\rm  d(d, n)}^3{\rm He}	^{\bf (*)}$ \hfil {\bf (5)}   &$ ^3{\rm He(n,p)t}$ \hfil {\bf (6)} &${\rm  t(d,n)}^4{\rm He}$ {\bf (7)} & $^3{\rm He(d, p)}^4{\rm He}  ^{\bf (*)}$ \hfil {\bf (8)} \\ \hline
$^3{\rm He}(\alpha,\gamma)^7{\rm Be}$ \hfil {\bf (9)} & ${\rm t(}\alpha,\gamma)^7{\rm Li}$ \hfil {\bf (10)}& $^7{\rm Be(n,p)}^7{\rm Li}$ \hfil {\bf (11)} &$^7{\rm Li(p,} \alpha )^4{\rm He}	^{\bf (*)}$ \hfil {\bf (12)}\\ \hline
\hline
\end{tabular} 
\caption{Nuclear reactions of greatest relevance for big bang nucleosynthesis, labelled from 1 through 12. Reactions measured with the Trojan Horse method are marked with a ${\bf (*)}$ symbol.}\label{rent} 
\end{center}
\end{table}  

The reaction rates are	calculated from the available low-energy cross sections  for reactions which are also a fundamental information for a number of other still unsolved astrophysical problems, e.g. the so called  ``lithium depletion" either in the Sun or in other galactic stars \break \cite{WM63,EC63}. Cross sections should be measured in the astrophysically relevant Gamow window \cite{iliadis,bertulanibook}, of the order of few hundreds keV. In the last decades these reactions have been widely studied and, in particular, great efforts have been devoted to their study by means of direct measurements at the relevant astrophysical energies, sometimes in underground laboratories  \cite{bonetti,casella} or improved detection systems.

For these extreme low energy cross section measurements and in several physical cases, new phenomena, such as the electron screening effect can no longer be neglected. 
This can significantly alter the low-energy cross section in direct experiments \cite{assenbaum87} due to the partial screening of nuclear charges by atomic electron clouds.
 For applications  to astrophysical plasmas one needs the ``bare nucleus" cross section, $\sigma_b$, especially because the screening in stellar conditions is much different from the one in laboratory. 
 
Moreover, for many of the relevant reactions, no direct experiments exist at astrophysical energies (mostly because of difficulties connected with the presence of the Coulomb barrier in charged particle induced reactions) and the cross section within the Gamow window has to be extrapolated from higher energy measurements. Alternative and challenging ways to obtain $\sigma_b$ for charged-particles at sub-Coulomb energies have been provided by indirect methods such as the Coulomb dissociation method \cite{BBR88,bertulani10} and the ANC (Asymptotic Normalization Coefficient) \cite{akram}. Among them, the Trojan-horse Method (THM) \cite{spitaleri99,spitaleri01} is particularly suited to investigate binary reactions induced at astrophysical energies by neutrons or charged particles by using appropriate three-body reactions. It allows one to avoid both Coulomb barrier suppression and electron screening effects, thus preventing the use of unreliable extrapolations.
 
An experimental program has been carried out during the last decade to apply the THM to study reactions of relevance for the SBBN (reactions marked in Table \ref{rent} with a $^{(*)}$). In the next sections, we will first discuss the available direct data for these 4 reactions in order to calculate their rate. In a subsequent section we will show the calculations of the reaction rates based also on the THM measurements of the cross sections $\sigma_b$. The THM  has been applied to several reactions in the past decade \cite{t08,lacon15,lacognata05,laco2011,leti2010,lamiab10,lamia,lamia12,romano,wen,pizzone05,pizzone09}, at the energies relevant for astrophysics, which usually are far below the Coulomb barrier, of the order of MeV's.
Many tests have been made to fully explore the potentiality  of the method and extend as much as possible its applications: the target/projectile break-up invariance \cite{musumarra}, the spectator invariance \cite{letizia,pizzone11,pizzone13} and the possible use of virtual neutron beams \cite{tuminoenam,gulino}.
Such studies are necessary, as the Trojan Horse method has become one of the major tools for the investigation of reactions
of astrophysical interest (for recent reviews see \cite{physicsnuclei}). Thus, the method can be regarded as a powerful indirect technique to get information about bare nucleus cross section for reactions of astrophysical interest, which  leads to new reaction rates determination. 

\begin{figure}[t]
\begin{center}
\includegraphics[width=105mm]{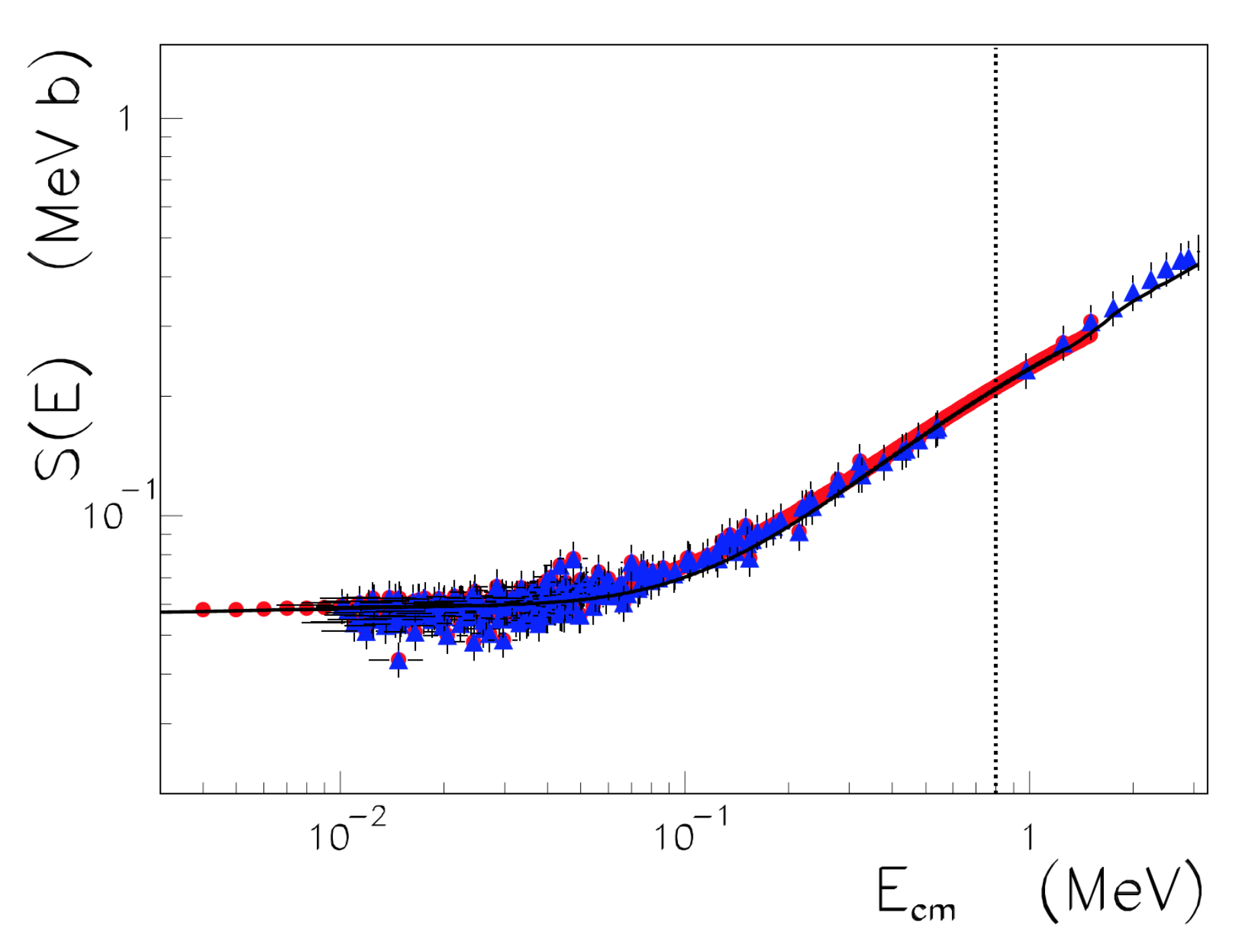}
\caption{Experimental data for the S-factor obtained for the reaction $^{2}$H(d,p)$^{3}$H with direct methods (blue solid triangles) and with the Trojan Horse method (red solid circles) data taken from \protect\cite{tumino11,tumino14}. The solid line is an R-matrix fit following described in section \ref{rmatsect}. The parameters for an equivalent polynomial fit are listed in Table  \ref{tabsf1}. The vertical dotted line marks the upper value of the energy range of interest for primordial nucleosynthesis.}\label{figddptsf}
\end{center}
\end{figure}

\section{S-factors and Reaction Rates}

The energy dependence of the bare nucleus cross sections is usually expressed in terms of the equation
\begin{equation}
 \sigma_b(E) = {S(E)\over E}\exp\left(-{2\pi\eta E}\right),
\label{se}
\end{equation}
where $S(E)$ is the astrophysical factor (or simply S-factor), $\eta=Z_iZ_je^2/\hbar v$ is the Sommerfeld parameter, with $Z_i$ the i-th nuclide charge, $v$ the relative velocity of the $ij$-pair  and $E=\mu v^2/2$ is the relative energy of $i+j$. 
The S-factor has a relatively weak dependence on the energy $E$, except when it is close to a resonance, where it is strongly peaked. 
Using standard nuclear physics units we write $2\pi \eta=b/\sqrt{E}$, where 
\begin{equation}
b=0.9898 Z_iZ_j\sqrt{A} \ {\rm MeV}^{1/2}, \label{b}
\end{equation} 
and $A$ is the reduced mass of $i+j$ in amu.  

The thermonuclear reaction rate at temperature $T$ is obtained from an average over the Maxwellian velocity distribution \cite{Fow67}
\begin{eqnarray}
R_{ij}=\frac{N_{i}N_{j}}{1+\delta_{ij}} \langle \sigma v \rangle = \frac{N_{i}N_{j}}{1+\delta_{ij}}\left(\frac{8}{\pi A}\right)^{\frac{1}{2}}\left(\frac{1}{k_BT}\right)^{\frac{3}{2}}
\int_{0}^{\infty}dES(E) \exp\left[-\left(\frac{E}{k_BT}+2\pi\eta(E)\right)\right], \label{rij}
\end{eqnarray}
where $\sigma$ is the fusion cross section, $v$ is the relative velocity of the $ij$-pair and $N_i$ is the number of nuclei of species $i$. 
The factor $1+\delta_{ij}$ in the denominator of Eq. \eqref{rij} corrects for the double-counting when $i=j$. 

We will express our reaction rates in the form $N_A \langle \sigma v \rangle$ (in units of cm$^3$ mol$^{-1}$ s$^{-1}$), where $N_A$ is the Avogadro number and $\langle \sigma v \rangle$ involves the integral in Eq. \eqref{rij} with the Maxwell distribution. For charged particles, a good accuracy (within 0.1\%) is reached using the integration limits between  $\max(0,E_0 - 5\Delta E)$ and $E_0+5\Delta E$, where in terms of  $T_9$ (temperature in units of $10^9$ K), the effective Gamow energy is given by
\begin{equation}
E_0=0.122(Z_i^2Z^2_jA)^{1/3}T_9^{2/3} \ {\rm MeV}, \label{gamow}
\end{equation}
and the Gamow energy window by
\begin{equation}
\Delta E=0.2368(Z_i^2Z^2_jA)^{1/6}T_9^{5/6} \ {\rm MeV}. \label{gamoww}
\end{equation}

\begin{figure}[tb]
\begin{center}\includegraphics[width=105mm]{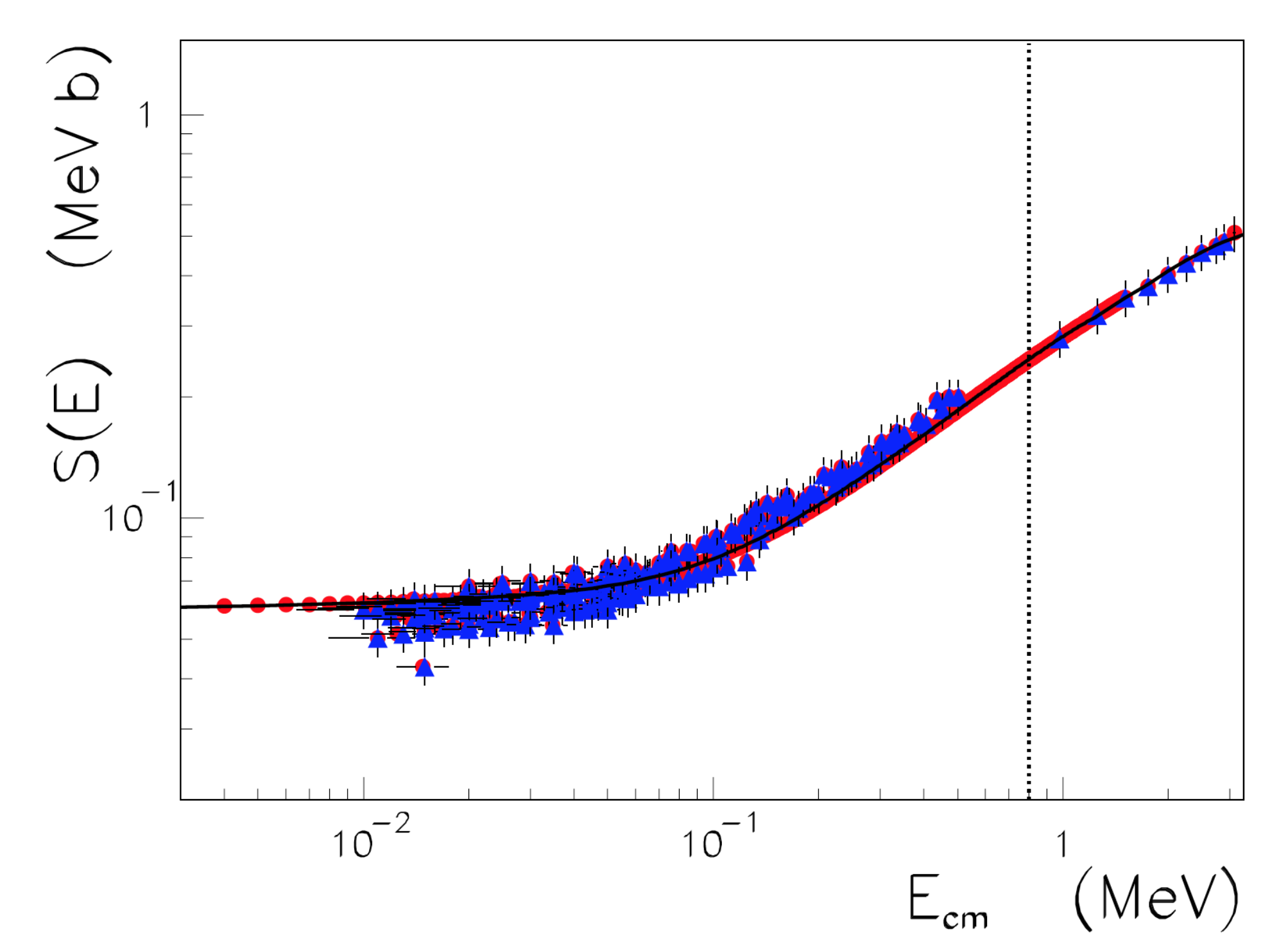}
\caption{Experimental data for the S-factor of the reaction $^2$H(d,n)$^3$He obtained with direct data (blue filled triangles) and with the Trojan Horse method (red  filled circles) taken from \protect\cite{tumino11}. The solid line is an R-matrix fit to both direct and indirect data following the general lines  described in section \ref{rmatsect}. The parameters for an equivalent polynomial fit are listed in Table  \ref{tabsf1}. The vertical dotted line marks the upper value of the energy range of interest for primordial nucleosynthesis.}
\label{figddn3hesf}
\end{center}

\end{figure}

\subsection{R-matrix fit}\label{rmatsect}

In contrast to polynomial fitting, the R-matrix method uses basic quantum mechanics as a guide for the data fitting and therefore is a preferable method. R-matrix fits are particularly useful in the presence of resonances. The energy dependence of the cross sections arises from matching the solutions of the  Schr\"odinger equation at a fixed  distance from the origin (channel radius). Cross sections and phase shifts are then reproduced with a small number of parameters (depending on the channel radius $a$), allowing for an extrapolation of the cross section down to astrophysical energies.
The channel radius divides the space into two regions: the internal region (with radius $a$), where nuclear forces are relevant, and the external region, where the interaction between the nuclei is governed by the Coulomb force. The R-matrix fits are usually weakly dependent on the channel radius. Matching at the channel radius leads to a number $N$ of S-matrix poles characterized by energy $E_\lambda$ and reduced width $\tilde{\gamma}_{\lambda}$. The
R-matrix at energy $E$ is defined as
\begin{equation}
R_{ij}(E)=\sum_{\lambda=1}^N {\tilde{\gamma}_{\lambda i} \tilde{\gamma}_{\lambda j} \over  E_\lambda -E}, \label{rmatfit}
\end{equation}
where the indices $i$ and $j$ refer to the reaction channels. These also involve total {momenta} $J$ and parities $\pi$. The reduced widths are directly proportional to the square of the solutions of the Schr\"odinger equation for the internal region calculated at the channel radius $a$.

From the R-matrix one deduces the collision matrix with help of which one can calculate the process of astrophysical interest, namely, radiative capture, transfer, elastic scattering, and rearrangement reactions.  The R-matrix formula in Eq. \eqref{rmatfit} can be used to fit both resonant and non-resonant reactions. For a non-resonant case one  uses a pole at large energies which simulates a background and yields an R-matrix almost independent of energy. A recent review of the R-matrix theory is found in Ref. \cite{DB10}. In our fits we use the multilevel, multichannel R-matrix public code, AZURE \cite{Azu10}. The code finds a best chi-square fit of the R-matrix parameters similar to Eq. \eqref{chi2fit}.

{For the 4 reactions marked with a (*) symbol in Table \ref{rent} we have performed a function fit to the R-matrix output, using direct and THM data (in the energy range where available, and using the direct data where not). The fit function was parametrized by a sum of polynomials and Breit-Wigner functions in the form 
\begin{equation}
S_{fit}(E)=\sum_{i=1}^6 b_i E^{i-1} +\sum_{j=1}^{n_R} {c_j\over (E-E_j)^2+\Gamma_j^2/4} \ \ \ \ \  ({\rm MeV \cdot b}),\label{sfit}
\end{equation}
where $n_R$ is the number of resonances in the fit, $E_j$ (in MeV) are the resonance energies and $\Gamma_j$ (in MeV) are their widths. 
We applied the ordinary $\chi^2$ statistics, defined by the minimization of
\begin{equation}
\chi^2 = \sum_i {\left[S_{exp}(E_i)-S_{fit}(E_i;b_1,c_1,E_{R1},\Gamma_1, \dots)\right]^2\over \sigma_i^2}, \label{chi2fit}
\end{equation}
where $S_{exp}(E_i)$ are the S-factors at energies $E_i$, $\sigma_i$ are the measured errors, and $S_{fit}$ given by Eq. \eqref{sfit}.  The fit function is then used in the calculation of the reaction rates using Eq. \eqref{rij}. }

\begin{figure}[t]
\begin{center}
\includegraphics[width=105mm]{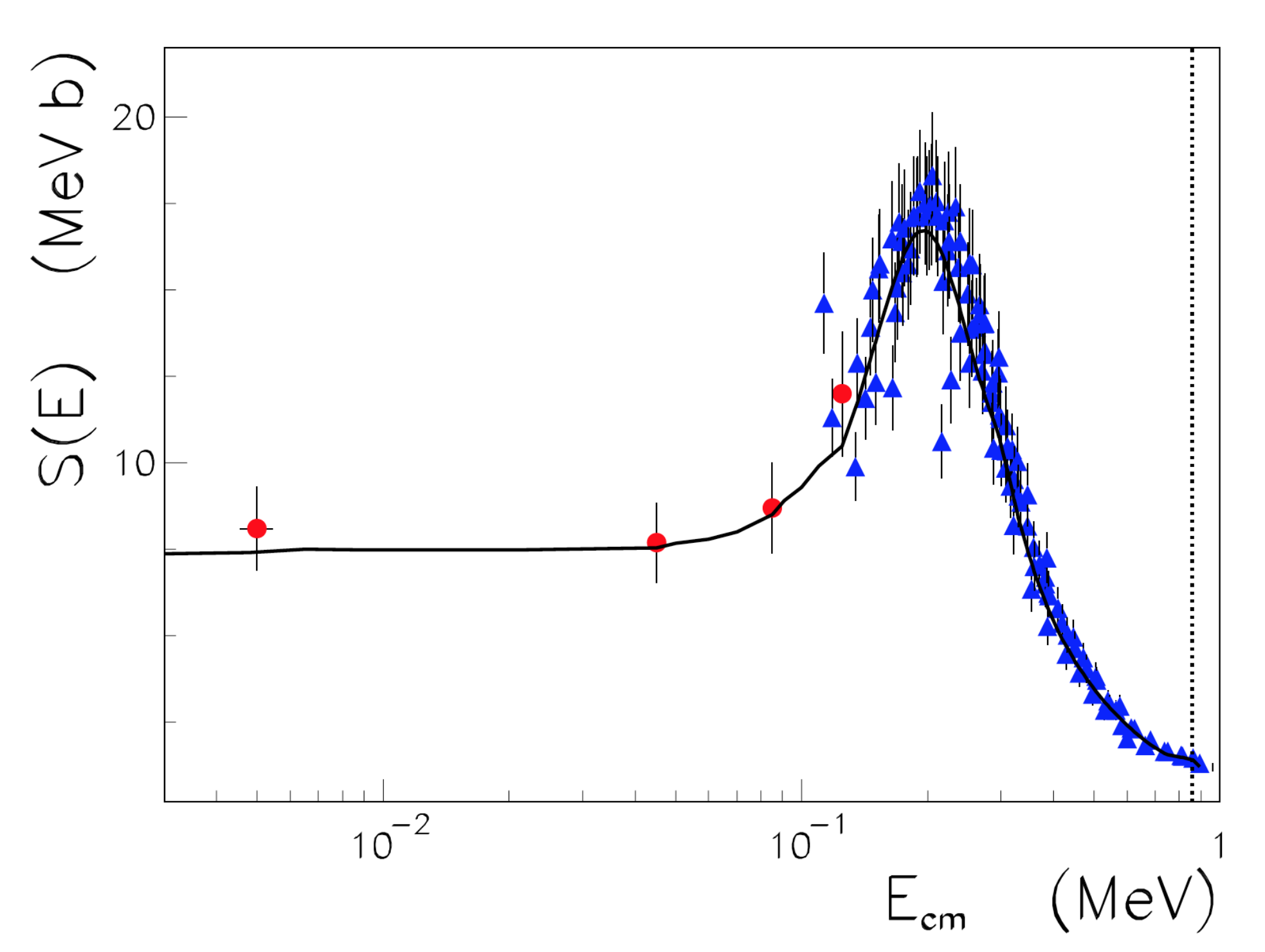}
\caption{Experimental data for the S-factor of the reaction $^3$He(d,p)$^4$He obtained with direct data (blue filled triangles) and with the Trojan Horse method (red filled circles) taken from \protect\cite{lacognata05}. The solid line is an R-matrix fit to both direct and indirect data following the general lines described in section \ref{rmatsect}. The parameters for an equivalent polynomial plus Breit-Wigner fit are listed in Table  \ref{tabsf2}. The vertical dotted line marks the upper value of the energy range of interest for primordial nucleosynthesis.}\label{fig3hedpasf}
\end{center}
\end{figure}

\section{BBN with Trojan Horse Data}
The reactions of interest for the SBBN cited in the introduction, i.e. $^7$Li(p,$\alpha$)$^4$He,  $^2$H(d,p)$^3$H, $^2$H(d,n)$^3$He, $^3$He(d,p)$^4$He, were studied by means of the THM in the energy range of interest and their measurements were performed in an experimental campaign which took place in the last decade \cite{aa,aa1,lacon15,lacognata05,laco2011,leti2010,lamiab10,lamia,lamia12,romano,wen,pizzone05,pizzone09}.
We will not go into the details of the THM because this is done elsewhere  \cite{spitaleri01} but we want to point out that the THM provides the bare S(E) factor, i.e., free of screening effects, for the reaction under investigation after studying an appropriate 3-body one in the quasi-free kinematical conditions.
The basic idea of the THM is to extract the cross section in the low-energy region of a two-body reaction with significant astrophysical impact:
  \begin{equation}
                        a+x\rightarrow c+C
  \label{binsubr1}
  \end{equation}
from a suitable three-body quasi-free (QF) reaction

\begin{equation}
a + b \to s + c + C.
\label{THM1}
\end{equation}
We therefore consider an interaction between the impinging nucleus and one of the clusters constituting the target (called participant x), while the residual nucleus, or spectator, does not participate in the reaction.
 In all the examined cases the extracted astrophysical S(E) factors were compared after the normalization procedure with those available from direct measurements and showed to be in fair agreement in the region where screening effects are negligible. The function fit parameters to the S(E) factors, obtained with Eq. \eqref{sfit} for the four reactions of relevance for BBN are listed in Tables \ref{tabsf1} and \ref{tabsf2} in units of MeV and barns. We will now review the available results for each reaction, taking into account the direct measurements available in literature as well as the new THM results mentioned above.

\begin{table}[htbp]
\vspace{0.0cm}
\centering
\begin{tabular}{|l|l|l|l|l|}
\hline
\hline
Parameter&$^{2}$H(d,p)$^{3}$H  
& $^{2}$H(d,n)$^{3}$He    
\\ \hline\hline
$b_1$ &$5.5325 \times 10^{-2}$    
&$5.8613\times 10^{-2}$ 
\\ \hline
$b_2$ &0.18293 
& 0.18101 
\\ \hline
$b_3$ &0.28256
& 0.44676
\\ \hline
$b_4$ &$-0.62121$    
&  $-0.8682$
\\ \hline
$b_5$ &0.44865    
& 0.61893
\\ \hline
$b_6$ &$-0.11278$    
& $-0.15675$
\\ \hline
\hline
\end{tabular} 
\caption{Table of  fit parameters (in Eq. \ref{sfit}) for the S-factors of the reactions $^{2}$H(d,p)$^{3}$H and $^{2}$H(d,n)$^{3}$He measured in TH experiments, as reported in the text. 
The coefficients $b_i$ are given in terms of MeV and barns.}\label{tabsf1} 
\end{table}    

\begin{table}[htbp]
\vspace{0.0cm}
\centering
\begin{tabular}{|l|l|l|l|l|}
\hline
\hline
Parameter&$^{3}$He(d,p)$^{4}$He  
& $^{7}$Li(p,$\alpha$)$^{4}$He    
\\ \hline\hline
$b_1$ &$1.7096$ 
&$-2.8141\times 10^{-2}$ 
\\ \hline
$b_2$ &$-20.121$ 
& $2.6584 \times 10^{-2}$ 
\\ \hline
$b_3$ &38.975
& $-2.7907 \times 10^{-2}$
\\ \hline
$b_4$ &$-20.406$ 
&  $-1.9457 \times 10^{-3}$
\\ \hline
$b_5$ &   \ \quad --
& $ 9.4651\times 10^{-4}$
\\ \hline
$b_6$ &   \ \quad -- 
& $-5.0471\times 10^{-4}$
\\ \hline
$c_1$ &$0.49562$ 
& 0.3198
\\ \hline
$E_{R1}$ &$0.24027$ 
& 2.5765
\\ \hline
$\Gamma_{R1}$ &$0.35011$ 
& 1.1579
\\ \hline
$c_2$ &  \quad --
& $9.7244 \times 10^{-2}$
\\ \hline
$E_{R2}$ &   \ \quad  --
& 5.0384
\\ \hline
$\Gamma_{R2}$ &   \quad  --
& 0.79323
\\ \hline
$c_3$ &   \quad  --   
& 0.40377
\\ \hline
$E_{R3}$ &   \ \quad  --  
& 6.0159
\\ \hline
$\Gamma_{R3}$ &   \ \quad  --  
& 1.8935
\\ \hline
$c_4$ &   \quad  --   
&1.9247
\\ \hline
$E_{R4}$ &   \ \quad  --    
& 8.0614
\\ \hline
$\Gamma_{R4}$ &   \ \quad  --    
& 4.0738
\\ \hline
\hline
\end{tabular} 
\caption{Table of  fit parameters (in Eq. \ref{sfit}) for the S-factors of the reactions $^{3}$He(d,p)$^{4}$He and $^{7}$Li(p,$\alpha$)$^{4}$He measured in TH experiments. 
The coefficients $b_i$ and $c_i$ are given in terms of MeV and barns. Energies and widths are in units of MeV.}\label{tabsf2} 
\end{table}

\begin{figure}[t]
\begin{center}
\includegraphics[width=105mm]{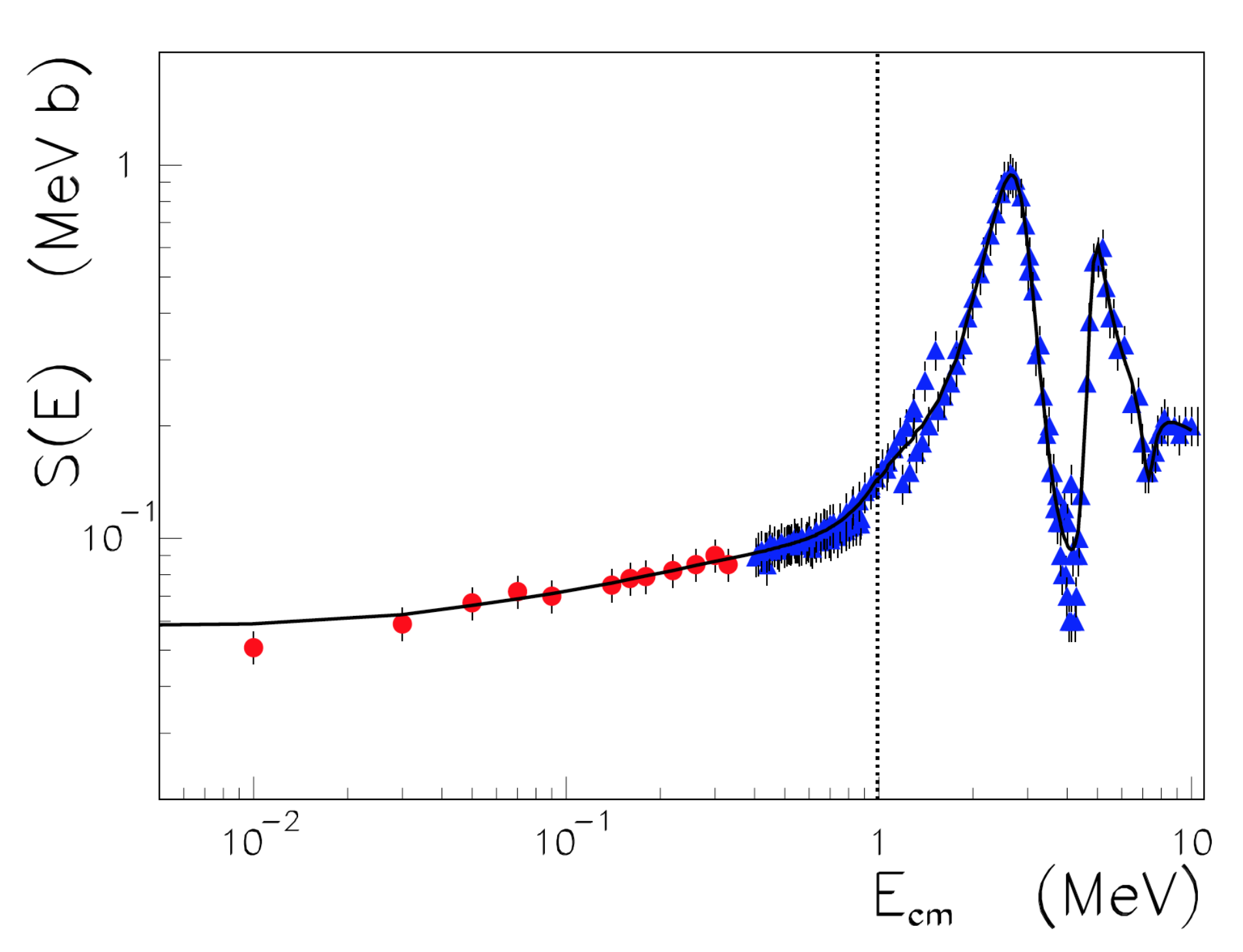}
\caption{Experimental S(E)-factor of the reaction $^7$Li(p,$ \alpha$)$^4$He obtained with direct data (blue filled triangles) and with the Trojan Horse method (red filled dots) taken from \protect\cite{lamia13}. The solid line is an R-matrix fit to both direct and indirect data following the general lines described in section \ref{rmatsect}. The parameters for an equivalent polynomial fit are listed in Table  \ref{tabsf2}. The vertical dotted line marks the upper value of the energy range of interest for primordial nucleosynthesis.}\label{fig7lipaasf}
\end{center}
\end{figure}

\subsection{$^2$H(d,p)$^3$H}

The d+d cross section has been extensively measured in laboratory for both the two \emph{mirror channels} $^{2}$H(d,p)$^{3}$H and $^{2}$H(d,n)$^{3}$He. Considering only results with a center-of-mass energy of interest for our purpose (i.e. around 1 MeV), the direct choice of data has been done accurately, selecting the newest and most reliable data sets, taking into account the possible presence of systematic errors. For $^{2}$H(d,p)$^{3}$H we chose the ones reported in \cite{Greife,krauss,McNeill,Schulte,BJ90,Ganeev,Arnold,Raiola2002,Booth,Davenport,VonEngel,Cook,Moffat,WangTieShan} and the most recent result from \cite{Leonard}. 
 
For the $^{2}$H(d,p)$^{3}$H the  data set of Ref. \cite{Greife} reaches down to a center-of-mass energy value of 1.62 keV, but this experiences a clear enhancement for very low energies because of the electron screening effect. Thus, in order to be used for astrophysical application, they need to be corrected for this effect. It is also noticeable that the energy range between 600 keV and 1 MeV is not covered by any data set, making it difficult to provide a reliable fit in the whole energy range.

The Trojan Horse experiment for this channel has been performed in two runs by measuring  the three-body reaction $^{2}$H($^{3}$He,pt)H. 
The data analysis, performed according to the THM prescriptions, allowed to measure the bare nucleus S-factor in the energy range from 2.6 keV up to 1.5 MeV, with a 5\% error (a full review is given in \cite{tumino11,tumino14}).

In Figure \ref{figddptsf} we show the data for the  S-factor  for the reaction $^{2}$H(d,p)$^{3}$H obtained with the Trojan Horse method (blue filled circles) and by different direct measurements (red  red circles). 
The solid line is an R-matrix fit to both direct and indirect data, as described in section \ref{rmatsect}. The parameters for an equivalent polynomial fit, using Eq. \eqref{sfit}, are listed in Table  \ref{tabsf1}.

\subsection{$^2$H(d,n)$^3$He}
The \emph{status of the art} before the THM measurement of the $^{2}$H(d,n)$^{3}$He is very similar to its mirror channel. The most relevant data sets are missing in the range between 600 keV and 1 MeV and in addition no experimental points in absolute units are present below 6 keV. For this reaction we have used for our fit to direct data \cite{Greife,krauss,McNeill,Schulte,BJ90,Leonard,Ganeev,Arnold,Raiola2002,Booth,Davidenko,Hofstee,Preston,Belov,Ying,Bystritsky2010}.

The bare nucleus S-factor has been obtained from the $^{2}$H($^{3}$He,n$^{3}$He)H by means of the THM \cite{tumino11}, and shown in fig. \ref{figddn3hesf}, 
with a 5\% experimental error on the whole data set, from 2.6 keV up to 1.5 MeV.
The data for the  S-factor  for the reaction $^2$H(d,n)$^3$He obtained with the Trojan Horse method (blue filled triangles) and by different direct measurements (red filled triangles) are shown in Figure \ref{figddn3hesf}. 
The solid line is an R-matrix fit  to the direct and THM data described in section \ref{rmatsect}. The parameters for an equivalent polynomial fit, using Eq. \eqref{sfit}, are listed in Table  \ref{tabsf1}.

\begin{figure}[H]
\begin{center}
\includegraphics[width=85mm]{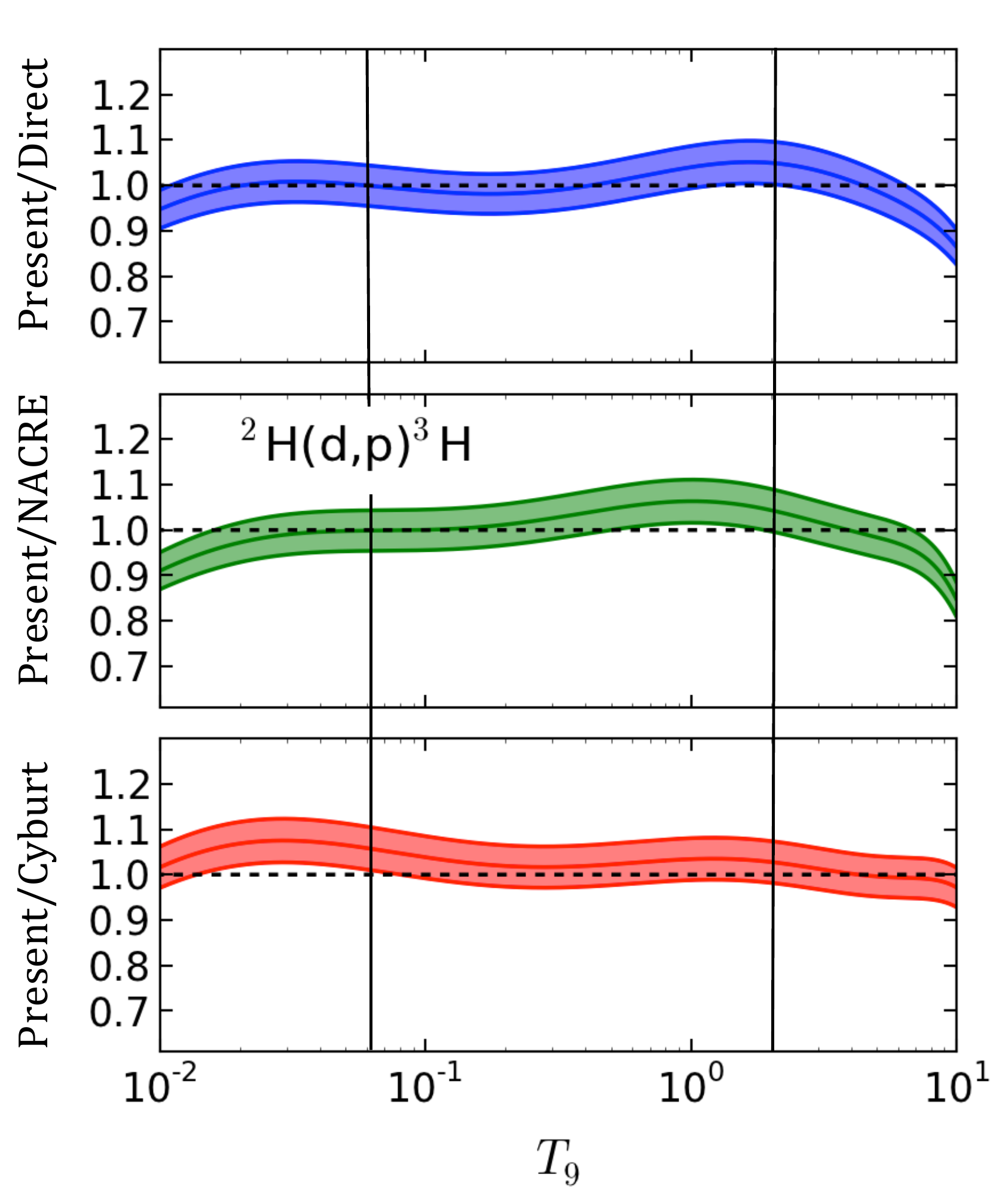}\includegraphics[width=80mm]{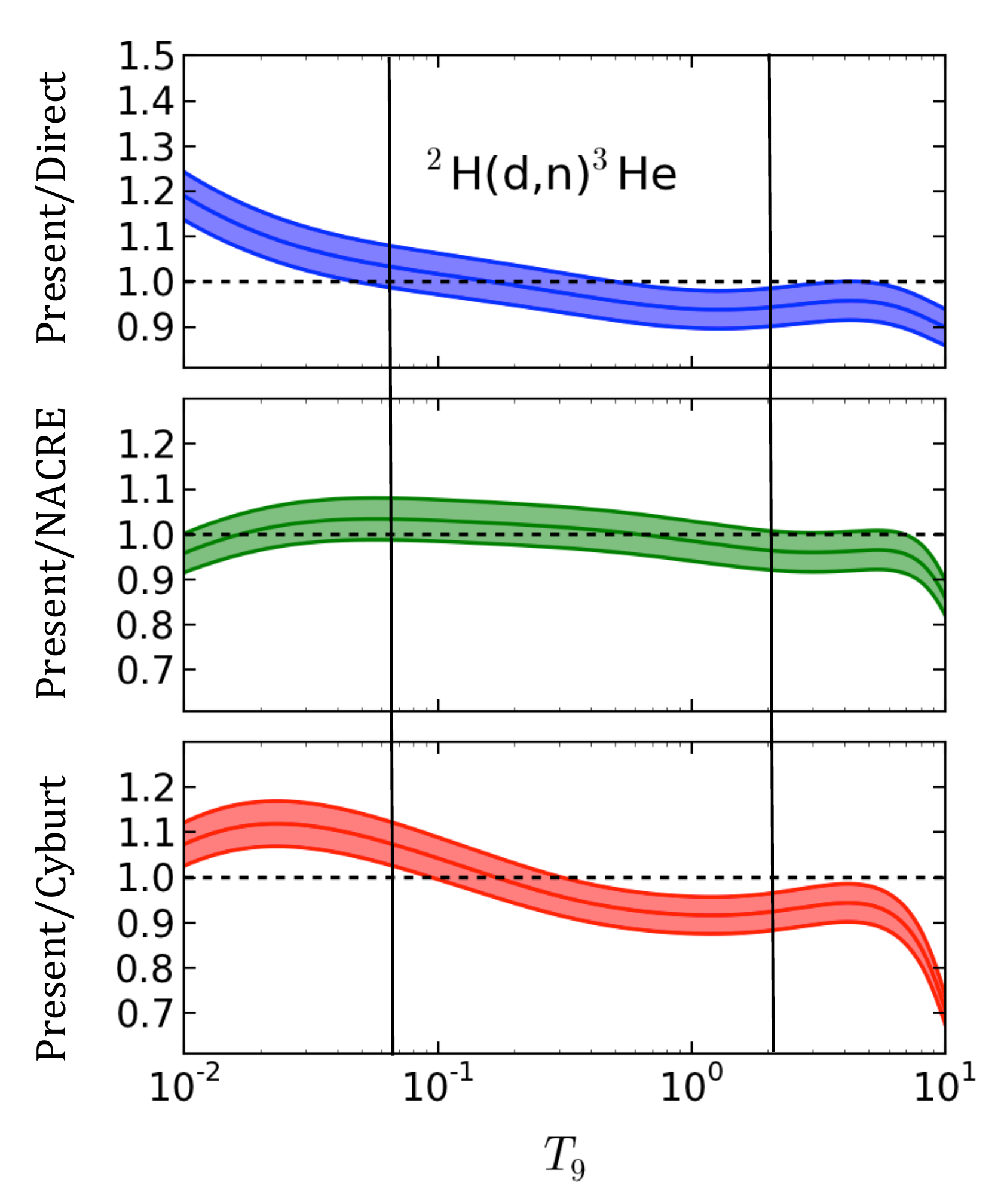}
\caption{{\it Left:} Ratio of the $^2$H(d,p)$^3$H reaction rates calculated using THM data to the one obtained from direct data fits (upper panel). The middle and lower panels are similar ratios using rates published in NACRE \cite{nacre} and Cyburt \cite{Cyb04}. {\it Right:} Same as in the left figure but for the  $^2$H(d,n)$^3$He reaction. The vertical lines represent the approximated lower and
upper temperature limits of interest for big bang nucleosynthesis.}\label{figrddpt}
\end{center}
\end{figure}

\subsection{$^3$He(d,p)$^4$He}

The bare-nucleus cross section for the $^3$He(d,p)$^4$He fusion reaction, at ultra-low energies, is of interest in pure and applied physics and was measured in the energy region of interest for astrophysics by means of several methods both indirect and direct \cite{aliotta,geist,krauss}. For the $^{3}$He(d,p)$^{4}$He we used the direct data from \cite{Engstler,krauss,Bonner,Zhicang,Geist,Moller,Erramli,Schroeder,Aliotta}. The THM experiment was perfomed by 
measuring the $^3$He($^6$Li,p$\alpha$)$^4$He reaction in quasi-free kinematics. The bare nucleus S(E)-factor was then extracted in the $0-1$ MeV energy range and fitted following Eq. \eqref{sfit}, as reported in \cite{lacognata05}. 
The S-factor  for the reaction $^3$He(d,p)$^4$He  is shown in figure \ref{fig3hedpasf} with red solid circles for THM data
 and blue filled triangles for the direct measurements. The solid line is an R-matrix fit  to the direct and THM data described in section \ref{rmatsect}. The parameters for an equivalent polynomial plus Breit-Wigner fit are listed in Table  \ref{tabsf2}. 

\subsection{$^7$Li(p,$ \alpha$)$^4$He}

Being the main channel of Li burning in astrophysical environments, this reaction is involved in the challenging scenarios of both stellar and primordial Li nucleosynthesis. In such case the discrepancy of about a factor of three between the predictions of SBBN  and the Li abundances observed in halo stars represents the well-known and still open ``lithium problem". A large number of possible explanations of this discrepancy have been proposed, from stellar phenomena, to non-standard Big Bang nucleosynthesis models. 

The $^7$Li(p,$ \alpha$)$^4$He reaction was extensively studied in the last 20 years both directly \break \cite{Engstler,Cruz} and indirectly \cite{lattuada01,aa,lamia13}, using the THM. For this reaction we used the data-sets from \cite{Schroeder,Ma,Cassagnou,Fiedler,Spinka,Rolfs,Harmon,Engstler92,Ciric,Spraker,ChulChuLee,Cruz}.   
The most recent data-set for the S-factor for this reaction, obtained with the Trojan Horse method after $d$ quasi-free breakup, are shown in Figure \ref{fig7lipaasf} \cite{lamia12} as red filled circles while the direct ones are reported as blue filled triangles. The solid line is an R-matrix fit to both direct and indirect data following the general lines described in section \ref{rmatsect}. The parameters for an equivalent polynomial fit are listed in Table  \ref{tabsf2}. The R-matrix fit is then used to calculate the reaction rate following Eq. \eqref{rij}, as for the other examined cases.

\begin{figure}[H]
\begin{center}
\includegraphics[width=75mm]{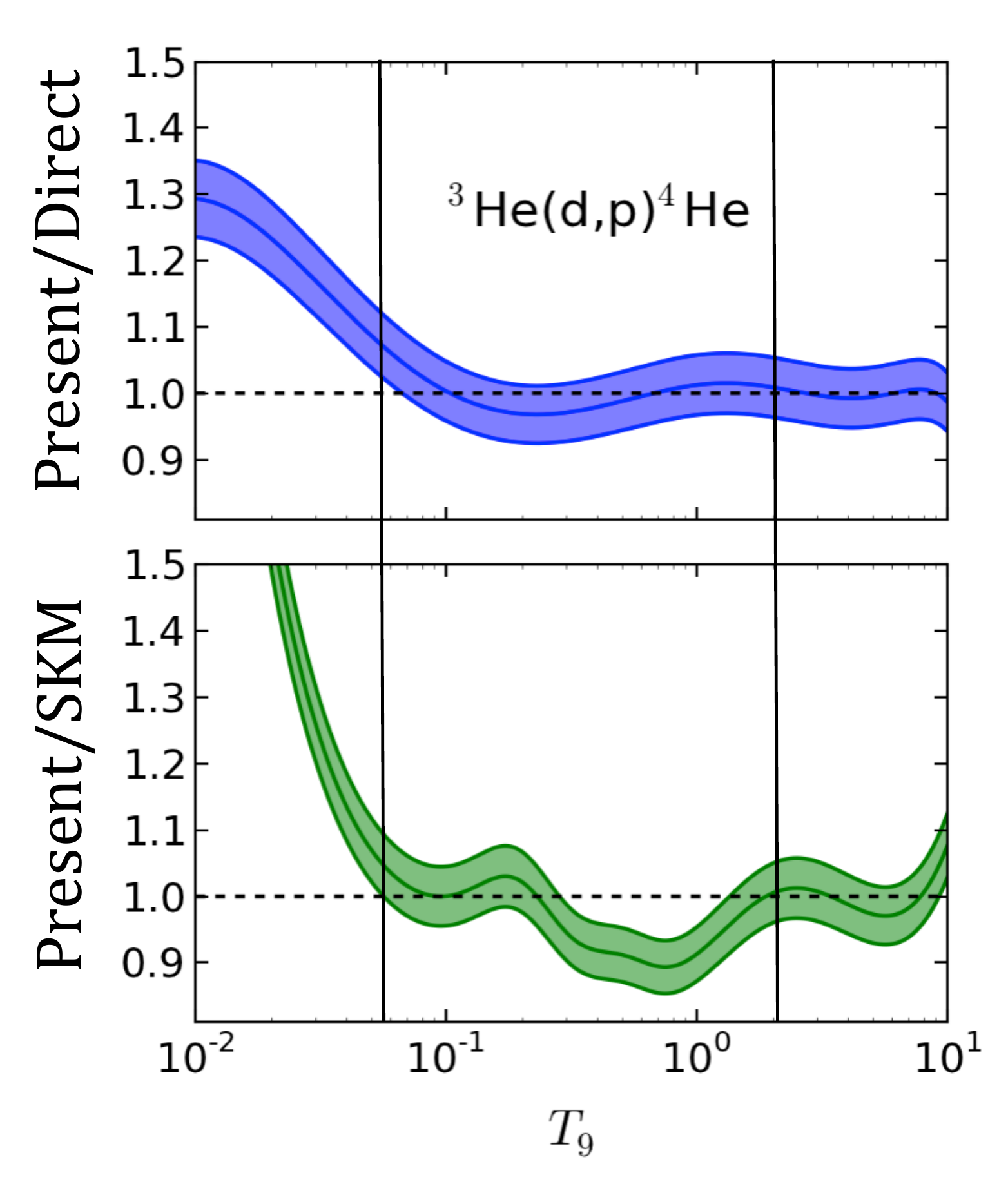}\includegraphics[width=75mm]{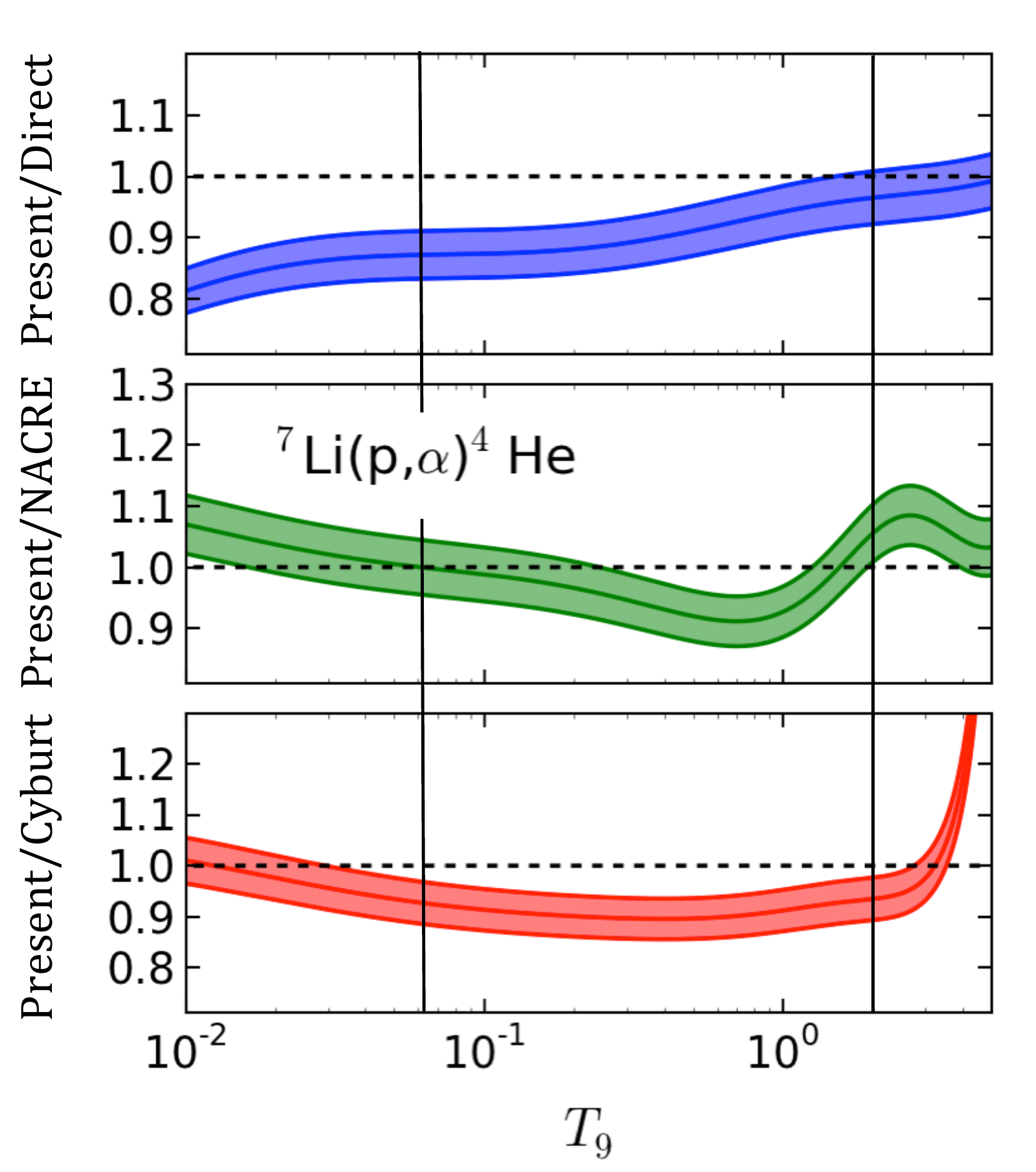}
\caption{{\it Left:} Ratio of the $^3$He(d,p)$^4$He reaction rates calculated using THM data to the one obtained from direct data fits (upper panel). The middle and lower panels are similar ratios using rates published in Smith, Kawano and Malaney \cite{SKM} (lower panel). {\it Right:} In the left figure, ratio of rates calculated using THM (as discussed in the text) to the one obtained with a fit to the direct data (without TH) of the S-factors (upper panel). The middle and lower panels are similar ratios with rates published in NACRE \cite{nacre} and Cyburt \cite{Cyb04}. The vertical lines represent the approximated lower and
upper temperature limits of interest for big bang nucleosynthesis.}\label{fig2}
\end{center}
\end{figure}

\subsection{Reaction rates with TH data}

The reaction rates for the the four reactions mentioned above (from a compilation of direct and THM data, as reported in the sections above) have been carried out numerically introducing the R-matrix results in Eq. \eqref{rij}. Thus, we fitted the rates with the parametrization displayed in Equation \eqref{NA}. This is the common procedure adopted in previous works (see, e.g., \cite{SKM,Cyb04,Alan12}). For the 4 reactions of interest, we have included the experimental errors from measurements, allowing us to evaluate the respective errors in the reaction rates. 
The numerical results are then fitted with the expression 
\begin{eqnarray} \label{NA}
N_{A}\left<\sigma v\right>=\exp\left [a_1 + a_2\ln T_9 + \frac{a_3}{T_9} + a_4 T_9^{-1/3} + a_5 T_9^{1/3}      
+ a_6T_9^{2/3} + a_7 T_9 + a_8T_9^{4/3} + a_9 T_9^{5/3} \right] , 
\end{eqnarray}
which incorporates the relevant temperature dependence of the reaction rates during the BBN.
The a$_i$ coefficients for the $^{2}$H(d,p)$^{3}$H and the $^{2}$H(d,n)$^{3}$He reactions are given for both  THM and direct measurements as well as for the direct ones (see next section for details) in Table \ref{parpar}, while the coefficients for the $^{3}$He(d,p)$^{4}$He and $^{7}$Li(p,$\alpha$)$^{4}$He reaction rate expression are given in Table \ref{parpar1}. 
The direct data were considered for energies above 100 keV for $^{3}$He(d,p)$^{4}$He and  $^{7}$Li(p,$\alpha$)$^{4}$He and for energies above 10 keV for $^{2}$H(d,p)$^{3}$H  and  $^{2}$H(d,n)$^{3}$He, in order to avoid the enhancement due to the electron screening in the direct data. 

\begin{table}[htbp]
\vspace{0.0cm}
\centering
\begin{tabular}{|l|l|l|l|l|}
\hline
\hline
$a_i$&$^{2}$H(d,p)$^{3}$H (present) &$^{2}$H(d,p)$^{3}$H  (direct)    & $^{2}$H(d,n)$^{3}$He (present)   & $^{2}$H(d,n)$^{3}$He  (direct)\\ \hline\hline
$a_1$ &14.996    &20.255 &16.1787 & 13.3209\\ \hline
$a_2$ & $-2.4127$ & $-0.63670$& $-1.9372$&$-2.9254$\\ \hline
$a_3$ &$2.8261 \times 10^{-3}$ &$7.7756\times 10^{-5}$ &  $2.0671\times 10^{-3}$&$ 4.0072\times 10^{-3}$ \\ \hline
$a_4$ &$-5.3256$    &$-4.2722$  & $-5.0226$&$-5.6687$ \\ \hline
$a_5$ & 6.6125    &$-1.0758$  & $5.7866$&$ 10.1787$ \\ \hline
$a_6$ &2.4656    & 2.3211  &  $-2.039\times 10^{-2}$ & 0.1550\\ \hline
$a_7$ &$-3.8702$    &$-1.3062$ & $ -0.7935$&  $-2.5764$\\ \hline
$a_8$ &1.6700    & 0.38274&0.2678 &$1.1967$   \\ \hline
$a_9$ &$-0.25851$    &$-5.0848\times10^{-2}$ & $-3.1586\times 10^{-2}$&$-0.1807$ \\ \hline
\hline
\end{tabular} 
\caption{Table with reaction rate parameters (appearing in Eq. \ref{NA}) for $^{2}$H(d,p)$^{3}$H and $^{2}$H(d,n)$^{3}$He evaluated from the present work and S-factors from direct measurements.}\label{parpar} 
\end{table}     
\begin{table}[htbp]
\vspace{0.0cm}
\centering
\begin{tabular}{|l|l|l|l|l|}
\hline
\hline
$a_i$&$^{3}$He(d,p)$^{4}$He (present) &$^{3}$He(d,p)$^{4}$He  (direct)    & $^{7}$Li(p,$\alpha$)$^{4}$He (present)   & $^{7}$Li(p,$\alpha$)$^{4}$He  (direct)\\ \hline\hline
$a_1$ &20.4005    &38.9078&17.6686 &17.5315 \\ \hline
$a_2$ & $1.3850$ & 5.9512& $-1.1549$ &  $-1.397$    \\ \hline
$a_3$ &$-1.2982 \times 10^{-2}$ &$-1.6061\times 10^{-2}$ &  $-4.4059\times 10^{-4}$&$6.9425\times 10^{-4}$ \\ \hline
$a_4$ &$-4.1193$    &$-2.1962$  & $-8.5485$&$-8.7921$ \\ \hline
$a_5$ & 12.2954    &$-20.5983$  & $4.6683$&$5.7430$ \\ \hline
$a_6$ &$-15.2114$    & 1.5636  &  $-0.7858$ &  $-2.4092$\\ \hline
$a_7$ &5.4147    &0.7040 & $-2.3208$&0.6434\\ \hline
$a_8$ &$-0.5048$    &$-0.1877$&2.0628 &$1.290$   \\ \hline
$a_9$ &$-4.3372\times10^{-2}$    &$2.9419\times10^{-2}$ & $-0.4747$&$-0.3467$ \\ \hline
\hline
\end{tabular} 
\caption{Table with reaction rate parameters (appearing in Eq. \ref{NA}) for $^{3}$He(d,p)$^{4}$He and $^{7}$Li(p,$\alpha$)$^{4}$He evaluated from present work and S-factors from direct measurements.}\label{parpar1} 
\end{table}     

The ratio between the reaction rates obtained with the THM with those from other compilations are shown in Figures \ref{figrddpt} and \ref{fig2}. In these figures, the comparison is made with reaction rates calculated from our own fit to existing direct reaction capture data, from the NACRE compilation \cite{nacre}, from the Smith, Kawano and Malaney compilation \cite{SKM}, and from the compilation by Cyburt \cite{Cyb04}. 
The error band is associated with the errors bars of the associated THM+direct S-factors (Figures \ref{figddptsf}-\ref{fig7lipaasf}). 

For all the cases we noticed that deviations of up to 20\% are obtained from previous compilations. The reaction rate for $^3$He(d,p)$^4$He process was not published in the NACRE compilation and in this case, we use the reaction rate fit as published in Ref. \cite{SKM}. Also, for the reaction $^7$Li(p,$\alpha$)$^4$H a large discrepancy with the reaction rate by Cyburt \cite{Cyb04} was found at temperatures above $T_9 \sim 4$.
 
\section{Application to big bang nucleosynthesis}

The SBBN is sensitive to certain parameters, including the baryon-to-photon ratio, the number of neutrino families, and the
neutron decay lifetime. We use the values $\eta = 6.16 \times 10^{-10}$ for the baryon-photon ratio, the  number of neutrino families $N_\nu = 3$, and the neutron lifetime $\tau_n = 878.5$ s, respectively (\cite{Stei07} and references therein). Our BBN abundances were calculated with a modified version of the standard BBN code derived from Refs. \cite{WFH67,Kaw68,Kaw92}.

Although BBN nucleosynthesis  can involve reactions up to the CNO cycle \cite{Alan12},   the  most important  reactions which can significantly affect the predictions of the abundances of the light elements are listed in Table \ref{rent}. The reaction rates involving Be, B, C, N and O isotopes were taken from Refs. \cite{Wag69,CF88,MF89,Wie89,Tho93}. For the remaining reactions we have used the compilations by SKM \cite{SKM}, NACRE \cite{nacre}, Cyburt \cite{Cyb04} and Descouvemont \cite{Des04} (and references mentioned therein). 
The data for the n(p,$\gamma$)d reaction was taken from the on-line ENDF database \cite{ENDF} - see also \cite{Cyb04,And06}. 

\begin{figure}[H]
\begin{center}
\includegraphics[width=105mm]{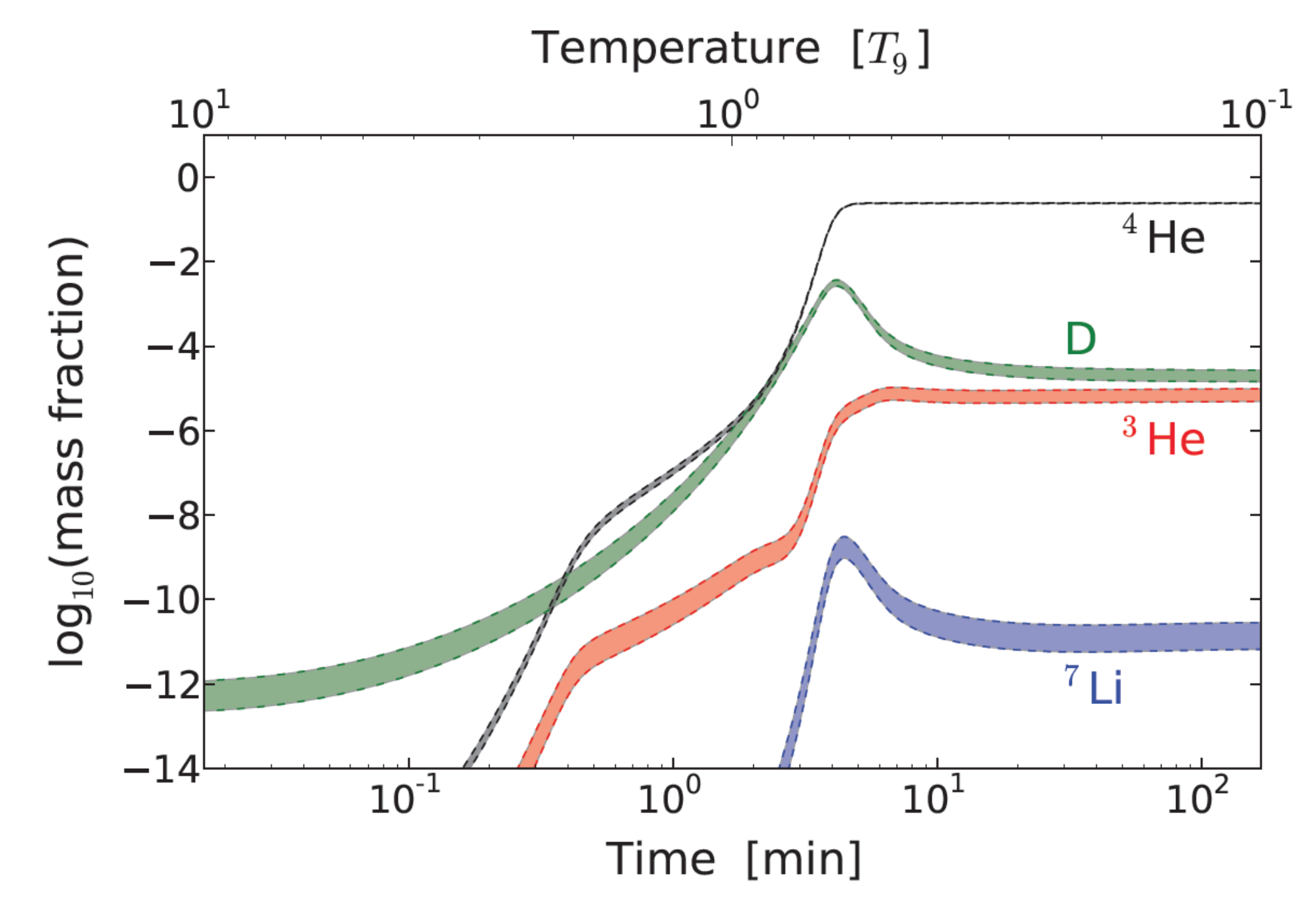}
\caption{Calculated BBN abundance of $^{3,4}$He, D and $^7$Li as a function of time and temperature. Black line represents $^4$He mass fraction, green the deuterium abundance, red the $^3$He abundance and blue the $^7$Li abundance. The error band represents the uncertainty in the THM measurements and their influence on the abundances.  }\label{fig5}
\end{center}
\end{figure}

In Figure \ref{fig5} we show the results for the abundances (mass fraction $Y_p$ for $^4$He) for $^2$H, $^3$He, $^4$He, $^6$Li and $^7$Li as a function of time  The uncertainties in the experimental nuclear data are reflected in the width of the predicted abundances. In Table \ref{tabbbn} the first column is the result obtained with our own fit to the world data from direct measurements. The second column is the impact of replacing the direct data for the  reaction $^{2}$H(d,p)$^{3}$H by those obtained with the reaction rate calculated in the present work. The subsequent columns are the same, but for the three other remaining  measurements. The column labelled ``all" uses all four reaction rates calculated in this paper. Finally, the last column lists the observed abundances. The uncertainties in the experimental data are reflected in the errors for the predicted abundances.   As expected, the abundance of $^4$He is barely affected by the new measurements. Also, some of the abundances are not affected by changes in reactions not directly related to the production or destruction of the element. Some appreciable changes in the abundances of d, $^3$He and $^7$Li are visible.

The mass fraction for $^4$He, $Y_p=0.2565 \pm 0.006$ (0.001 statistical and 0.005 systematic), was  obtained from the observation of low-metallicity extragalactic HII regions \cite{YT10} . The mean deuterium abundance is $\left< ({\rm D/H})\right > = (2.82 \pm 0.26) \times 10^{-5}$, which is equivalent to $\Omega_b h^2 \ ({\rm BBN}) = 0.0213 \pm 0.0013$ \cite{Mea06}. This average agrees within error bars with $\Omega_b h^2 \ ({\rm CMB})	=	0.02273	\pm	0.00062$	obtained from the analysis of {WMAP5 \cite{Dun09} (see also Ref. \cite{Stei10,Pet08}).}
The $^3$He abundance is adopted from Ref. \cite{BRB02} as  a lower limit to the primordial abundance. The lithium abundance arises from observations of stars which provide a sample of the ``lithium plateau" \cite{Sbo10}.
In Figure \ref{fig5} it is reported the calculated abundance for $^{3,4}$He, $^2$H and $^7$Li as a function of time and temperature for the BBN. The band represents the uncertainty derived from the measurements discussed above for each element. 
The present work gives an exhaustive and updated review of the rate reaction evaluation for some of the relevant reactions for nuclear astrophysics (arising both from direct and indirect methods). We can point out that the discrepancy calculated-to-observed for $^3$He and $^7$Li (see Table \ref{tabbbn}) is still evident, as seen in many other investigations \cite{Stei07}) and it seems to not be due to nuclear reaction rates uncertainties.

\begin{table}[htbp]
\vspace{0.0cm}
\centering
\caption{\label{tab:Table1} BBN predictions using different set of data (see text) compared with observations. (a) The mass fraction for $^4$He, $Y_p=0.2565 \pm 0.006$ (0.001 statistical and 0.005 systematic), is from Ref. \cite{YT10}.  (b) The mean deuterium abundance is the mean average $\left< ({\rm D/H})\right > = (2.82 \pm 0.26) \times 10^{-5}$, which is equivalent to $\Omega_b h^2 \ ({\rm BBN}) = 0.0213 \pm 0.0013$ \cite{Mea06}.  (c) The $^3$He abundances are adopted from Ref. \cite{BRB02} as  a lower limit to the primordial abundance. (d) The lithium abundance arises from observations of stars which provide a sample of the ``lithium plateau" \cite{Sbo10}. D/H is in units of 10$^{-5}$, $^3$He/H in 10$^{-6}$ and Li/H in 10$^{-10}$.
}
\begin{tabular}{|l|c|c|c|c|c|c|c|}
\hline
\hline
\small
{\small Yields} &{\small  Direct} &  {\small  $^{2}$H(d,p)$^{3}$H    }         &{\small d(d,n)$^3$He}&{\small $^3$He(d,p)$\alpha$}&{\small $^7$Li(p,$\alpha$)$^4$He }&all & {\small Observed}\\  \hline
{\small  $Y_p$ }&{\small 0.2486}&{\small 0.2485$^{+0.001}_{-0.001}$}&{\small 0.2485$^{+0.000}_{-0.000}$}&{\small 0.2486$^{+0.000}_{-0.000}$}&{\small 0.2486$^{+0.000}_{-0.000}$}&{\small 0.2485$^{+0.001}_{-0.002}$}&{\small $0.256\pm 0.006^{(a)}$ }\\ \hline
{\small D/H }
& 2.645&2.621$^{+0.079}_{-0.046}$&{\small 2.718$^{+0.077}_{-0.036}$}&{\small 2.645$^{+0.002}_{-0.007}$}&{\small 2.645$^{+0.000}_{-0.000}$}&{\small 2.692$^{+0.177}_{-0.070}$}&{\small $2.82\pm 0.26^{(b)}$} \\ \hline
{\small ${^3}$He/H }
 &9.748&9.778$^{+0.216}_{-0.076}$&{\small 9.722$^{+0.052}_{-0.092}$}&9.599$^{+0.050}_{-0.003}$&9.748$^{+0.000}_{-0.000}$&9.441$^{+0.511}_{-0.466}$&$\geq 11.\pm 2.^{(c)}$\\ \hline
${^7}$Li/H 
&4.460&4.460$^{+0.001}_{-0.001}$&4.470$^{+0.010}_{-0.006}$&4.441$^{+0.190}_{-0.088}$&4.701$^{+0.119}_{-0.082}$&4.683$^{+0.335}_{-0.292}$&$1.58\pm 0.31^{(d)} $ \\ \hline

\hline
\end{tabular} 
\vspace{0.0cm}
\label{tabbbn}
\end{table}

\section{Discussion and Conclusion}
The reaction rates of 4 of the main reactions of the BBN network in the temperature range  (0.001$<$T$_9<$10), namely,  $^{2}$H(d,p)$^{3}$H, d(d,n)$^3$He, $^3$He(d,p)$^4$, $^7$Li(p,$\alpha$)$^4$He, have been calculated numerically including the recent THM measurements. { The uncertainties of experimental data for direct and THM data have been fully included. The extension of the same methodology to the other reactions forming the BBN reaction network will be examined in a forthcoming paper.}
 The parameters of each reaction rates as given in Eq. \ref{NA} are reported  in Tables \ref{parpar} and \ref{parpar1}. The obtained reaction rates are compared with the some of the most commonly used compilations found in the literature. An updated compilation of direct data for the $^{2}$H(d,p)$^{3}$H, d(d,n)$^3$He, $^3$He(d,p)$\alpha$, $^7$Li(p,$\alpha$)$^4$He reactions has also been made, and relative expressions for the reaction rate are also given. The reaction rates calculated in the present work are used to calculate the BBN abundance for $^{3,4}$He, D and $^7$Li. The obtained abundances are in agreement, within the experimental errors, with those obtained using the compilation of direct reaction rates. Moreover, a comparison of our predictions with the observations for primordial abundance of $^{3,4}$He, D and $^7$Li show an agreement for $^{3,4}$He and  D, while showing a relevant discrepancy for $^7$Li.  
The present results show the power of THM as a tool for exploring charged particle induced reactions at the energies typical of BBN. From Table \ref{tabbbn} we can see that the primordial abundances calculated using the present reaction rates, agree within the uncertainties with the predictions arising from direct data. The comparison between predicted values and  observations clearly confirms the discrepancy for $^7$Li  abundance which is still under debate.

\
\bigskip

This work was supported by the Italian Ministry of the University under Grant No. RBFR082838 and by the Italian Ministry of University MIUR under the grant ''LNS Astrofisica Nucleare (fondi premiali)''. This work was partially supported by the US-DOE grants DE-FG02-08ER41533 and DE-FG02-10ER41706. A.M. acknowledges the support by  US Department of
Energy under Grant Nos. DE-FG02-93ER40773, DE-FG52-
09NA29467, and DE-SC0004958.


\bigskip

\begin{thebibliography}{}
\bibitem[Aliotta et al. 1999]{Aliotta} M. Aliotta et al. 2001, Nucl. Phys. A 790, 690  
\bibitem[Aliotta et al. 2000]{aliotta} M. L. Aliotta, et al. 2000,  Eur. Phys. J. A9, 435
\bibitem[Ando et al. 2006]{And06} S. Ando et al. 2006, Phys. Rev. C 74, 025809
\bibitem[Angulo et al. 1999]{nacre} C. Angulo et al. 1999, Nucl. Phys. A656, 3
\bibitem[Arnold et al. 1954]{Arnold} W.B Arnold et al. 1954, Phys. Rev. 483, 93 
\bibitem[Assembaum et al. 1987]{assenbaum87} H.J. Assenbaum et al. 1987, Zeitschrift fŸr Physik A, 327, 461
\bibitem[Azuma et al. 2010]{Azu10} R. E. Azuma et al. 2010, Phys. Rev. C 81, 045805 
\bibitem[Bania et al. 2002]{BRB02}T. M. Bania, R. T. Rood, D. S. Balser 2002, Nature 415, 54 
\bibitem[Baur et al. 1986]{BBR88} G. Baur, C.A. Bertulani \& H Rebel 1986, Nucl. Phys. A458, 188 
\bibitem[Belov et al. 1990]{Belov} A.S. Belov et al. 1990, Nuovo Cimento A  1647, 103
\bibitem[Bertulani 2013]{bertulanibook} C.A. Bertulani 2013,  ``Nuclei in the Cosmos'', World Scientific    
\bibitem[Bertulani \& Gade 2010]{bertulani10} C.A. Bertulani \& A. Gade 2010, Phys. Rep. 485, 195   
\bibitem[Bonetti et al. 1999]{bonetti} R. Bonetti et al. 1999, Phys. Rev. Lett. 82, 5205 
\bibitem[Bonner et al. 1952]{Bonner}  T. W. Bonner et al. 1952, Phys. Rev. 88, 473 
\bibitem[Booth et al. 1956]{Booth} D.L Booth et al. 1956, Proc. Physical Society (London)  A 265, 69  
\bibitem[Bystritsky et al. 2010]{Bystritsky2010} V.M. Bystritsky et al. 2010, Izv. Rossiiskoi Akademi Nauk, 563, 74 
\bibitem[Casella et al. 2002]{casella} C. Casella et al. 2002, Nucl. Phys. A706, 203 
\bibitem[Cassagnou et al. 1962]{Cassagnou} Y. Cassagnou et al. 1962, Nucl. Phys. 449, 33
\bibitem[Caughlan \& Fowler 1993]{CF88}G.R. Caughlan and W.A. Fowler 1988, Atomic Data and Nuclear Data Tables 40, 283 
\bibitem[Ciric et al. 1976]{Ciric} D.M. Ciric  1976, Rev. of Science Research 115, 6 
\bibitem[Coc et al. 2012]{Alan12} A. Coc, S. Goriely, Y. Xu, M. Saimpert, and E. Vangioni 2012,  Astrophys. J.  744, 18 
\bibitem[Cook et al. 1953]{Cook} C.F. Cook et al. 1953, Phys. Rev. 785, 89 
\bibitem[Cruz et al. 2009]{Cruz} J. Cruz et al. 2009, Nucl. Instr. and Meth. in Phys. Research B 478, 267 
\bibitem[Cyburt  2004]{Cyb04} R.H. Cyburt 2004, Phys. Rev. D 70, 023505 
\bibitem[Davenport et al. 1953]{Davenport} P.A Davenport et al. 1953, Proc. Royal Society (London). A 66, 213  
\bibitem[Davidenko et al. 1957]{Davidenko} V.A. Davidenko et al. 1957,  J. Nucl. Energy 258, 2 
\bibitem[Descouvemont  \& Baye 2010]{DB10} P. Descouvemont  and D. Baye 2010, Rep. Pro. Phys. 73, 036301 
\bibitem[Descouvemont et al. 2004]{Des04} P. Descouvemont, A. Adahchour, C. Angulo, A. Coc, and E. Vangioni-Flam 2004, At. Data Nucl. Data Tables {88}, 203 
\bibitem[Dunkley et al. 2009]{Dun09} J. Dunkley et al. 2009, Astrophys. J. Suppl. 180, 306 
\bibitem[ENDF]{ENDF} ENDF/B-VI, Online database at the NNDC Online Data Service, http://www.nndc.bnl.gov.
\bibitem[Engstler et al. 1988]{Engstler} S. Engstler et al. 1988, Phys. Lett. B, 202, 179
\bibitem[Engstler et al. 1992]{Engstler92} S. Engstler et al. 1992, Phys. Lett. B 20, 279 
\bibitem[Erramli et al. 2005]{Erramli} H. Erramli et al. 2005, Physical and Chemical News (Morocco) 67, 23 
\bibitem[Ezer \& Cameron 1963]{EC63} D. Ezer and A.G.W. Cameron 1963, Icarus 1, 422
\bibitem[Fiedler et al. 1967]{Fiedler} O. Fiedler et al. 1967, Nucl. Phys. A 513, 96 
\bibitem[Fields \& Sarkar 2006]{Fie06}B. D. Fields and S. Sarkar 2006, J. Phys. G33, 220 
\bibitem[Fowler, Caughlan \& Zimmerman 1967]{Fow67} W.A. Fowler, G.R.  Caughlan, and B.A. Zimmerman 1967, Ann. Rev. Astron. Astrophys.  5, 525 
\bibitem[Ganeev et al. 1957]{Ganeev} A.S. Ganeev 1957, Suppl. of Sov. Atom. Journal, 5, 26
\bibitem[Geist et al. 1999]{geist} W. H. Geist, C. R. Brune, H. J. Karwowski, E. J. Ludwig, K. D. Veal, and G. M. Hale 1999, Phys. Rev. C 60, 054003 
\bibitem[Geist et al. 1999]{Geist} W.H. Geist et al. 1999, Phys. Rev. C60, 054003
\bibitem[Greife et al. 1995]{Greife} U. Greife et al. 1995, Z. Phys. 351, 107
\bibitem[Gulino et al. 2010]{gulino}{M. Gulino et al. 2010, Jour. of Phys.  {37}, 125105}
\bibitem[Harmon  1989]{Harmon} J.F. Harmon 1989, Nucl. Instr. and Meth. in Phys. Research B 507, 40
\bibitem[Hofstee et al. 2001]{Hofstee} M.A. Hofstee et al. 2001, Nucl. Phys. A688, 527   
\bibitem[Iliadis 2007]{iliadis}C. Iliadis 2007, ``Nuclear Physics of Stars'', Wiley 
\bibitem[Israelian 2012]{Isr12} G. Israelian 2012, Nature 489, 37 
\bibitem[Izotov \& Thuan 2010]{YT10} Y.I.Izotov and T.X.Thuan 2010, Ap. J. Lett. 710, L67 
\bibitem[Jarmie et al. 1990]{BJ90} N. Jarmie and R.E. Brown 1990, Phys. Rev. C 41, 1391  
\bibitem[Kawano 1988]{Kaw68} L. Kawano 1988, FERMILAB Report No. PUB-88/34-A (unpublished).
\bibitem[Kawano 1992]{Kaw92} L. Kawano 1992, NASA STI/Recon Technical Report N {92}, 25163
\bibitem[Kolb \& Turner 1990]{KT90} E.W. Kolb and M.S. Turner 1990, ``The Early Universe", Addison-Wesley
\bibitem[Komatsu 2011]{Kom11} E. Komatsu et al. 2011, Ap. J. Sup. 192, 18 
\bibitem[Krauss et al. 1987]{krauss}  A. Krauss, H. W. Becker, H. P. Trautvetter, C. Rolfs, and K. Brand 1987, Nucl. Phys. A465, 150 
\bibitem[La Cognata et al. 2005]{lacognata05}{M. La Cognata et al. 2005, Phys. Rev. C { 72}, 065802}
\bibitem[La Cognata et al. 2007]{lacon15}{M. La Cognata et al. 2007, Phys. Rev. C  { 76}, 065804  }
\bibitem[La Cognata et al. 2008]{lacognata08} M. La Cognata et al. 2008,  Phys. Rev. Lett. { 101}, 152501
\bibitem[La Cognata et al. 2011]{laco2011}{M. La Cognata et al. 2011, Ap. J. L. 739, L54}
\bibitem[Lamia et al. 2007]{lamiab10}{L. Lamia et al. 2007, Nucl. Phys.  {A 787}, 309c}
\bibitem[Lamia et al. 2012]{lamia}{L. Lamia, et al. 2012, J. Phys. G 39, 015106}
\bibitem[Lamia et al. 2012a]{lamia12}{L. Lamia,M. La Cognata, C. Spitaleri, B. Irgaziev, R.G. Pizzone 2012, Phys. Rev. C 85, 025805}
\bibitem[Lamia et al. 2013]{lamia13}{L. Lamia et al. 2013, A. \& A. 541, 158}
\bibitem[Lattuada et al. 2001]{lattuada01}{M. Lattuada et al. 2001, Ap. J. { 562}, 1076}
\bibitem[Lee  1969]{ChulChuLee} Chul Chu Lee 1969, J. of Korean Phys. Soc. 1, 2
\bibitem[Leonard et al. 2006]{Leonard} D.S. Leonard et al. 2006, Phys. Rev. C 73, 045801
\bibitem[Malaney \& Fowler 1989]{MF89}R.A. Malaney and W.A. Fowler 1989, Astrophys. J. 345, L5 
\bibitem[Mani et al. 1964]{Ma} G.S. Mani et al. 1964, Nucl. Phys. 588, 60 
\bibitem[McNeill et al. 1951]{McNeill} K.G. McNeill et al. 1951, Phys. Rev. 602, 81  
\bibitem[Moeller et al. 1980]{Moller} W. Moller et al. 1980, Nucl. Instr. and Meth. in Phys. Research 111, 168  
\bibitem[Moffat et al. 1952]{Moffat} J. Moffat et al. 1952, Proc. Royal Society (London). A 220, 212  
\bibitem[Mukhamezhanov et al. 2008]{akram} A. Mukhamezhanov et al. 2008, Phys. Rev. C {78}, 0158042008 
\bibitem[Musumarra et al. 2001]{musumarra} A. Musumarra et al. 2001, Phys. Rev. C { 64}, 068801  
\bibitem[O'Meara et al. 1999]{Mea06} J.M. O'Meara, S. Burles, J.X. Prochaska, and G.E. Prochter 2006,  Ap. J. 649, L61
\bibitem[Olive, Steigman \& Walker 2000]{OSW00} K. A. Olive, G. Steigman, and T. P. Walker 2000, Phys. Rep. 333, 389 
\bibitem[Pettini et al. 2008]{Pet08}M. Pettini, B. J. Zych, M. T. Murphy, A. Lewis, C. C. Steidel 2008, MNRAS 391, 1499 
\bibitem[Pizzone et al. 2003]{aa}R.G. Pizzone et al. 2003, A. \& A.  { 398}, 423   
\bibitem[Pizzone et al. 2005]{aa1}{R.G. Pizzone et al. 2005, A. \& A. { 438}, 779}
\bibitem[Pizzone et al. 2005b]{pizzone05} {R.G. Pizzone et al. 2005, Phys. Rev. C { 71}, 058801}
\bibitem[Pizzone et al. 2009]{pizzone09} {R.G. Pizzone et al. 2009, Phys. Rev. C {80}, 025807}
\bibitem[Pizzone et al. 2011]{pizzone11}R.G. Pizzone et al. 2011, Phys. Rev. C, 83,  045801 
\bibitem[Pizzone et al. 2013]{pizzone13}R.G. Pizzone et al. 2013, Phys. Rev. C 87, 025805 
\bibitem[Preston et al. 1954]{Preston} G. Preston et al. 1954, Proc. Royal Society (London) A 206, 226 
\bibitem[Raiola et al. 2002]{Raiola2002} F. Raiola et al. 2002, Eur. Phys. J. A13, 377
\bibitem[Rolfs et al. 1999]{Rolfs} C. Rolfs et al. 1986, Nucl. Phys. A  179, 455 
\bibitem[Romano et al. 2006]{romano} S. Romano et al. 2006, Eur. Phys. J. A {27}, 221 
\bibitem[Sbordone et al. 2010]{Sbo10}L. Sbordone,  et al. 2010,  Astron. Astro. 522, 26
\bibitem[Schroeder et al. 1989]{Schroeder} U. Schroeder et al. 1989, Nucl. Instr. and Meth. in Phys. Research B 466, 40 
\bibitem[Schulte et al. 1972]{Schulte} R.L. Schulte 1972, Nucl. Phys. A 192 
\bibitem[Sergi et al. 2010]{leti2010}{M.L. Sergi  et al.2010, Phys. Rev. { C 82}, 032801}
\bibitem[Smith et al. 1993]{SKM} M.S. Smith, L.H. Kawano and R.A. Malaney 1993, Ap. J. 85, 219 
\bibitem[Spinka et al. 1971]{Spinka} H. Spinka et al. 1971, Nucl. Phys.  A  1, 164   
\bibitem[Spitaleri et al. 1999]{spitaleri99}{C. Spitaleri et al. 2003, Nucl. Phys. A 719, 99c }
\bibitem[Spitaleri  et al. 2001]{spitaleri01}{C. Spitaleri et al. 2001, Phys.  Rev. C { 63}, 055801}
\bibitem[Spitaleri et al. 2011]{physicsnuclei}{C. Spitaleri et al. 2011, Physics of At. Nucleus 74, 1725}
\bibitem[Spraker et al. 1999]{Spraker} M. Spraker et al. 2000, Phys. Rev. C 015802, 61  
\bibitem[Steigman  2010]{Stei10} G.Steigman 2010, 11th Symposium on Nuclei in the Cosmos, NIC XI July 19-23, Heidelberg, Germany 
\bibitem[Steigman 2007]{Stei07}  G. Steigman 2007, Ann. Rev. Nucl. Part. Sci. 57, 463 
\bibitem[Thomas et al. 1993]{Tho93}D. Thomas, D.N. Schramm, K.A. Olive, and B.D. Fields 1993, Astrophys. J. 406, 569 
\bibitem[TieShan 2007]{WangTieShan} W. Tie-Shan 2007, Chin. Phys. Lett. 3103, 24 
\bibitem[Tumino et al. 2005]{tuminoenam}{A. Tumino et al. 2005, Eur. Phys. J.  A { 25}, 649 }
\bibitem[Tumino et al. 2006]{letizia} A. Tumino et al. 2006, Eur. Phys. J. A { 27} Supplement 1, 243 
\bibitem[Tumino et al. 2008]{t08}{A. Tumino et al. 2008, Phys. Rev. C { 78}, 064001}
\bibitem[Tumino et al. 2011]{tumino11}{A. Tumino et al. 2011, Phys. Lett. B  700 (2), 111}
\bibitem[Tumino et al. 2014]{tumino14}{A. Tumino et al. 2014, in press on Ap. J.}
\bibitem[Von Engel et al. 1961]{VonEngel} A. Von Engel et al. 1961, Proc. Royal Society (London).  A 445, 264  
\bibitem[Wagoner  1969]{Wag69} R.V. Wagoner 1969, Ap. J. Suppl. 162,  247 
\bibitem[Wagoner Fowler \& Hoyle 1967]{WFH67}R. Wagoner, W.A. Fowler, and F. Hoyle 1967, Astrophys. J. {148}, 3
\bibitem[Wen et al. 2008]{wen} Q. Wen et al. 2008, Phys. Rev. C { 78}, 035805
\bibitem[Weymann \& Moore 1963]{WM63} R. Weymann and E. Moore 1963,  Ap. J. 137, 552
\bibitem[Wiescher et al. 1989]{Wie89} M. Wiescher, R. Steininger  and F. Kappeler 1989,  Astrophys. J. 344, 464 
\bibitem[Ying et al. 1973]{Ying} N. Ying et al. 1973, Nucl. Phys. A 481, 206
\bibitem[Zhicang et al. 1977]{Zhicang} Li Zhicang et al. 1977, Atomic Energy Science and Tecnology, 229, 11 
\end{thebibliography}
\end{document}